\documentclass[9pt,a4paper,twoside]{rho-class/rho}
\usepackage[english]{babel}

\usepackage{algorithm}
\usepackage{algpseudocode}

\usepackage{listings}
\definecolor{codegreen}{rgb}{0,0.6,0}
\definecolor{codegray}{rgb}{0.5,0.5,0.5}
\definecolor{codepurple}{rgb}{0.58,0,0.82}
\definecolor{backcolour}{rgb}{0.95,0.95,0.92}

\lstdefinestyle{mystyle}{
    backgroundcolor=\color{backcolour},   
    keywordstyle=\color{orange},
    stringstyle=\color{codepurple},
    basicstyle=\ttfamily\footnotesize,
    breakatwhitespace=false,         
    breaklines=true,                 
    captionpos=t,                    
    keepspaces=true,                 
    numbersep=5pt,                  
    showspaces=false,                
    showstringspaces=false,
    showtabs=false,                  
    tabsize=2
}

\lstdefinelanguage{JavaScript}{
  keywords={typeof, new, true, false, catch, let, function, return, null, catch, switch, var, if, in, while, do, else, case, break},
  keywordstyle=\color{blue}\bfseries,
  ndkeywords={class, export, boolean, throw, implements, import, this},
  ndkeywordstyle=\color{darkgray}\bfseries,
  identifierstyle=\color{black},
  sensitive=false,
  comment=[l]{//},
  morecomment=[s]{/*}{*/},
  commentstyle=\color{purple}\ttfamily,
  stringstyle=\color{red}\ttfamily,
  morestring=[b]',
  morestring=[b]"
}

\definecolor{delim}{RGB}{20,105,176}
\definecolor{numb}{RGB}{106, 109, 32}
\definecolor{string}{rgb}{0.64,0.08,0.08}

\lstdefinelanguage{JSON}{
    frame=single,
    rulecolor=\color{black},
    showspaces=false,
    showtabs=false,
    breaklines=true,
    postbreak=\raisebox{0ex}[0ex][0ex]{\ensuremath{\color{gray}\hookrightarrow\space}},
    breakatwhitespace=true,
    basicstyle=\ttfamily\tiny,
    upquote=true,
    morestring=[b]",
    stringstyle=\color{string},
    literate=
     *{0}{{{\color{numb}0}}}{1}
      {1}{{{\color{numb}1}}}{1}
      {2}{{{\color{numb}2}}}{1}
      {3}{{{\color{numb}3}}}{1}
      {4}{{{\color{numb}4}}}{1}
      {5}{{{\color{numb}5}}}{1}
      {6}{{{\color{numb}6}}}{1}
      {7}{{{\color{numb}7}}}{1}
      {8}{{{\color{numb}8}}}{1}
      {9}{{{\color{numb}9}}}{1}
      {\{}{{{\color{delim}{\{}}}}{1}
      {\}}{{{\color{delim}{\}}}}}{1}
      {[}{{{\color{delim}{[}}}}{1}
      {]}{{{\color{delim}{]}}}}{1},
}
\usepackage{silence}

\setbool{rho-abstract}{true} 
\setbool{corres-info}{true} 

\title{CyberNFTs: Conceptualizing a decentralized and reward-driven intrusion detection system with ML}

\author[1]{Synim Selimi}
\author[1,$\dagger$]{Blerim Rexha}
\author[2]{Kamer Vishi}

\affil[1]{Department of Computer Engineering, University of Prishtina, Kodra e Diellit, p.n., 10000 Prishtina, Kosovo}
\affil[2]{Department of Informatics, University of Oslo, Gaustadalléen 23B, N-0373 Oslo, Norway}
\affil[$\dagger$]{Corresponding author}

\dates{To cite this work, please refer to the original published article as follows: Selimi, S., Rexha, B. and Vishi, K. (2022) CyberNFTs: Conceptualizing a decentralized and reward-driven intrusion detection system with ML, {\it International Journal of Metadata, Semantics and Ontologies}, Vol. 22, No. 1, pp.117\textendash 138.}

\leadauthor{Selimi et al.}

\corres{Blerim Rexha -}
\email{blerim.rexha@uni-pr.edu}
\doi{\url{https://doi.org/10.1504/IJICS.2023.133385}}

\license{\textbf{Note:} This is the author's version of the original publication in the International Journal of Information and Computer Security (ISSN: 1744-1765 for print, 1744-1773 for online).}

\begin{abstract}
    The rapid evolution of the Internet, particularly the emergence of Web3, has transformed the ways people interact and share data. Web3, although still not well defined, is thought to be a return to the decentralization of corporations' power over user data. Despite the obsolescence of the idea of building systems to detect and prevent cyber intrusions, this is still a topic of interest. This paper proposes a novel conceptual approach for implementing decentralized collaborative intrusion detection networks (CIDN) through a proof-of-concept. The study employs an analytical and comparative methodology, examining the synergy between cutting-edge Web3 technologies and information security. The proposed model incorporates blockchain concepts, cyber non-fungible token (cyberNFT) rewards, machine learning algorithms, and publish/subscribe architectures. Finally, the paper discusses the strengths and limitations of the proposed system, offering insights into the potential of decentralized cybersecurity models.

\end{abstract}


\begin{document}
	
    \maketitle
    \thispagestyle{firststyle}


\section{Introduction}

    \rhostart{T}he Internet and with it the ways of interaction between people have changed with increased momentum in recent years. The developments of each technological era have been summarized and classified continuously by the research community. Between the 90s and 05s, Web 1.0 was created to be used mainly as a communal resource, not individually, with open standards, decentralized use and mainly informative content. Web 2.0 was the big leap towards becoming a fully commercialized information gathering/sharing medium, enabling virtual interaction between users and the independent creation of online content, and as a consequence centralizing its impact and services in the hands of a few major corporations. For too long, Web3 (also known initially as Web 3.0) has been thought to be a return to the decentralization of power that corporations have over user data, which together with their semantic representation, cryptocurrencies, non-fungible tokens (NFTs), decentralization of organizations (DAO), finance (DeFi), artificial intelligence, 3D Graphics and metaverse tries to rebuild the way we understand the world of the Internet and the integral role that the Internet has in the functioning of contemporary society \cite{chohan2022web}.
    
    Decentralized finance is an example, although atypical, of the weight and influence that Web3 and blockchain have today in determining the technological future and the interaction of the legacy banking world versus a world of completely decentralized financial institutions. DeFi is undoubtedly a proof of concept for blockchain scalability that has already been proliferated quite well \cite{zheng2017overview}. The introduction and rise of a non-fungible token (NFTs) and tokenomics \cite{bao2022non} as a byproduct of DeFi in the form of tokenized cybercurrency also raises many questions regarding unexplored fields of application that decentralization has yet to visit. As explained in \cite{zheng2017overview} and \cite{aljaroodi2019blockchain}, the applicability of decentralization ideas in other fields and industries has been researched and implemented quite a lot so far, which makes the long-standing effort to use concepts of decentralization in cyber security or information security even more appealing.
    
    On the other hand, the ubiquity of the Internet and digitalization keeps the need for security mechanisms and cyber protection always relevant. Cyber intrusion detection and prevention systems (IDPS) are among the earliest legacy software solutions to implement such protections. Despite the maturity and age of this idea, it seems that it still remains a topic of research due to advancing cyberattacks, as it has been proven to be increasingly effective in commercial cyber defense, especially recently with the integration of machine learning (ML) algorithms. Even with these advancements, according to the UK's annual cyber security report for 2022, the growth of attacks and their impact on businesses and non-governmental organizations is continuous and uninterrupted, also illustrated by Figure 1. According to the report in \cite{gov2022cyber} only in the United Kingdom the damage caused by these cyber attacks and intrusions costs medium and large businesses an average of £20,000 per year.

\begin{figure}[ht]
\centerline{\includegraphics[width=1\columnwidth]{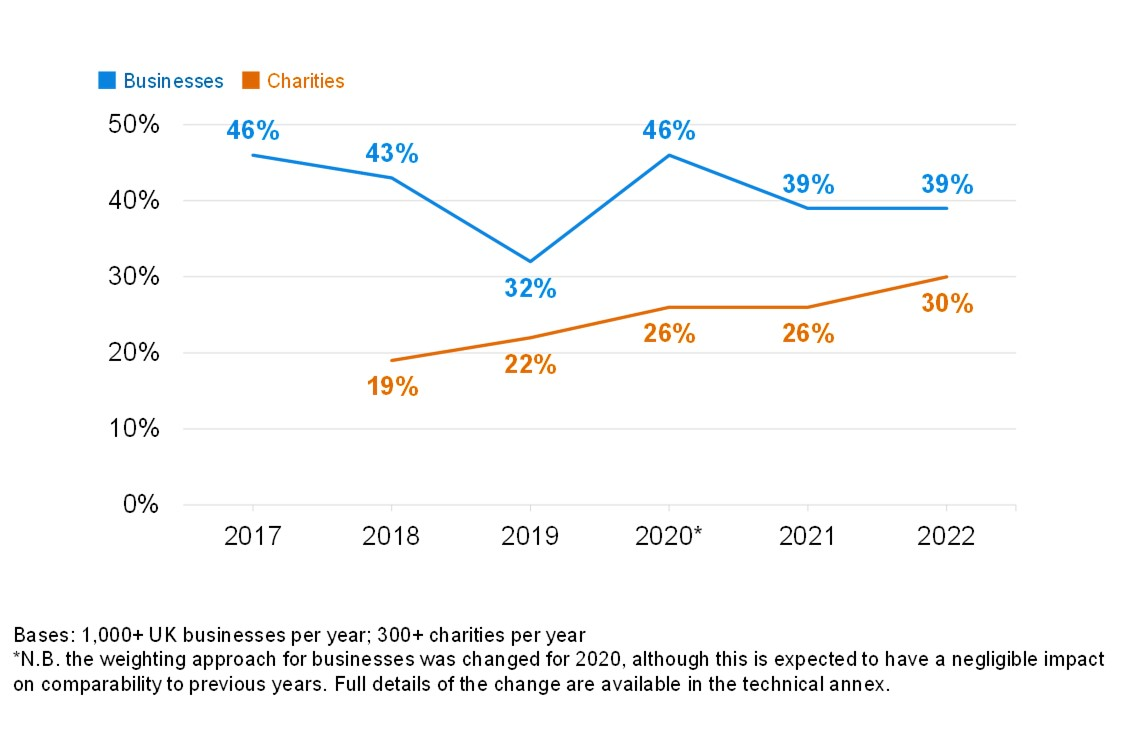}}
\centering
\caption{Proportion of businesses and charities reporting cyber intrusions according to UK annual report 2022 \cite{gov2022cyber}}
\label{fig:annual_report_2022}
\end{figure}

The analysis of the interaction between these latest state-of-the-art ML and blockchain technologies in the field of cyber security broadens the prism from which we see the usefulness of these technologies. This implementation and study clearly examines and demonstrates the limits of automation capabilities and autonomous systems for the detection and prevention of security intrusions. By simulating such an autonomous system, applicability challenges and differences can then be identified alongside existing, outdated, and conventional systems used today in centralized or semi-centralized on-premise and cloud architectures.

\subsection{Defining the problem}

Intrusion detection systems have had a continuous development since the early theoretical models \cite{denning1987intrusion} although not at the pace and momentum of other technologies such as the Internet, DeFi and the like. This may be due to the increased complexity that IDS systems represent, due to the depletion of field research in behavior pattern recognition (e.g. in \cite{biggio2014security}), due to the choice to invest in preventive ways or more cost-effective ways of protection, because of the decision to depend on cyber security experts instead of automated solutions and so on. All this makes IDS, CIDN and in general the field of automated cyber security architectures rather not researched enough in relation to the wider literature and new technologies, although there is research within narrow scopes and features such as pattern recognition, design and implementation of mobile agents, parametric classification of network traffic, etc.

Due to the still-to-be-defined nature and extremely broad scope of Web3, the effort to integrate Web3 features with modern cyber intrusion detection systems as a proof of concept is reduced and focused on only a few key Web3 features: decentralization, distributed operation, independent operation and gradual improvement in relation to the amount of data and/or the amount of time.

As presented in Figure 2, the conceptual architecture of the decentralized intrusion detection system will:
\begin{itemize}
    \item monitor local network traffic
    \item use ML models for classification
    \item classify and store traffic in a decentralized manner on the blockchain
    \item accentuate and reward the intrusion detection peer with a unique NFT
\end{itemize}

\begin{figure}[ht]
\centerline{\includegraphics[width=1\columnwidth]{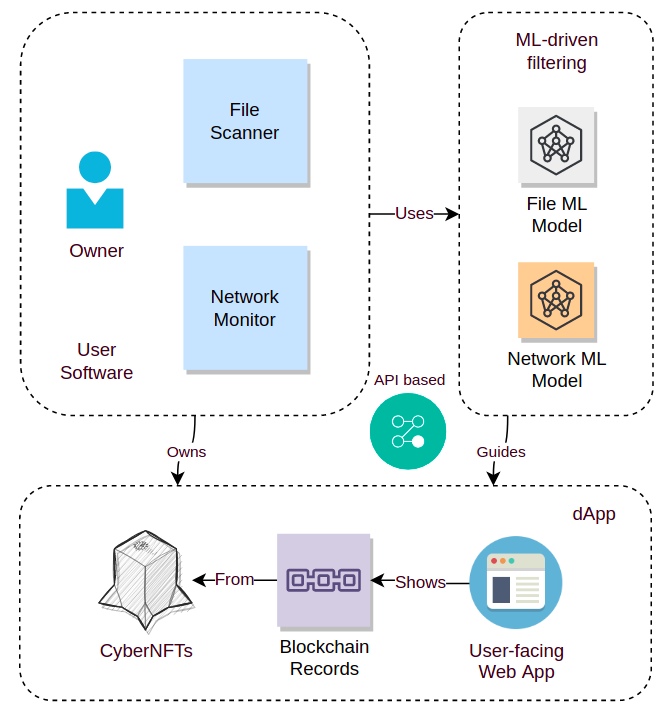}}
\centering
\caption{The conceptual architecture of the system and essential components}
\label{fig:conceptual_architecture}
\end{figure}

Due to the conceptual nature of this paper, no experimental analysis is included. Instead, this paper presents a theoretical exploration of the proposed approach and provides a proof-of-concept to demonstrate the potential of the approach, its advantages and limitations. This study aims to answer the following questions:
\begin{itemize}
    \item \textit{How is information decentralization and ML technologies applied to cyber intrusion detection systems?}
    \item \textit{What are the main challenges in developing a decentralized and collaborative intrusion detection network?}
\end{itemize}

The research methodology as well as the discussion of the problem are divided into four main parts. The first part includes analyzing how state-of-the-art IDSs work, with special emphasis on their integration with blockchain and ML. The second part involves the development and integration of these software solutions into a CIDN. The third part includes testing and evaluating the way this CIDN network works as a case study. The last part includes the analysis and comparison of the operation of this system and its features against existing ones.
\vspace{-0.3cm}
\section{Breakthrough of ML and collaborative IDS } \label{sec:datamanagementoverview}

The first models proposed for the design and implementation of intrusion detection systems in \cite{denning1987intrusion} are still quite in use and these statistical techniques of analyzing the behavior of computer users are still relevant to date, with slightly more refined and advanced levels of approximation. These models are primarily based on the hypothesis that security breaches and vulnerability exploitation can be detected by monitoring the system's audit/evidence data for abnormal patterns of system usage or behavior. By building profiles of the behavior of the actors in relation to the objects in metric and statistical models and through the recognition of the rules for extracting the nature of the behaviors from the audit data, it is possible to discover abnormal behaviors and risks.

While there are many models, architectures and implementations of cyber intrusion detection systems, there is still no known practical implementation, existing software or definitive patented architecture to implement and integrate IDS with blockchain technology, machine learning algorithms (ML) and unique ownership credit systems in mind. In the following section we will mention and discuss the latest research in each of the separate disciplines involved in this integration as well as the ongoing efforts to perform a holistic integration.

In \cite{disha2022performance}, the authors performed a comprehensive analysis of machine learning models in relation to intrusion detection systems such as, Decision Tree (DT), Gradient Boosting Tree (GBT), Multilayer Perceptron (MLP), AdaBoost, Long-Short Term Memory (LSTM), and Gated Recurrent Unit (GRU). Furthermore, the authors applied Gini Impurity-based Weighted Random Forest (GIWRF) as a feature selection technique and binary classification for intrusion detection systems. This research is used as a knowledge basis in our machine learning integration attempts. 

Moreover, \cite{biggio2014security} and \cite{borkar2017survey} discuss the detection of anomalies through data mining to extract datasets with signatures of anomalies and then in \cite{osken2019intrusion} and \cite{buczak2016survey} their classification using deep learning. Despite all these multidisciplinary improvements, the essential nature of recognizing user behaviors for anomaly detection remains the most effective way of building an IDS.

The classifiers used in \cite{firdausi2010analysis} such as kNN, Naïve Bayes, J48 Decision Tree, Support Vector Machine (SVM) and Multilayer Perceptron Neural Network (MlP) have achieved accuracy of 96.8\% and precision of 97.3\%. Such classification quotients for network traffic analysis are quite promising, therefore we'll be using them in this paper without focusing on improving precision, since the proof of concept focuses mainly on the architectural whole as an integration and not the construction as a process.

Regarding blockchain and technologies for decentralizing functionality, there is a sufficient body of theoretical, applied, and scientific knowledge that summarizes the features of blockchain architecture and compares the common consensus algorithms used in different blockchains. These are discussed in more detail and in a summarized manner in \cite{zheng2017overview}, while the forms of their application and the problems and different obstacles it can solve, are mentioned quite extensively in \cite{aljaroodi2019blockchain}. Blockchain technology offers some attractive features, also in the sense of protecting user's privacy and data integrity, consequently there is a natural tendency to shift applications towards blockchain technologies. \cite{neziri2022assuring} presents an approach to leverage blockchain technologies for electronic voting and thus assuring not only the anonymity and privacy of user votes but also its immutability. Where as in \cite{dervishi2022transactions} is presented a solution to preserve the user privacy using the web of trust concept. 

One step further to enhance the detection capabilities of an IDS is the development of cooperative intrusion detection networks (or cooperative IDS), which allow IDS nodes to exchange data with each other \cite{meng2018when}. Driven by this, there is theoretical and summarized research like \cite{kumar2020distributed} in relation to hybrid IDS and blockchain systems on stable platforms such as cloud infrastructure, which fall in direct relation and within our framework of discussion. Also \cite{kolokotronis2021blockchain}, focusing more on the aspect of distributed operation, proposes the use of a CIDN-based blockchain with consensus protocols and a form of chain of trust to protect the integrity of shared information between CIDN nodes to increase their availability and ensure their cooperation by preventing insider attacks.

In \cite{ajayi2019consortium} and \cite{ajayi2021blockchain} a solution is proposed that leverages the distributed blockchain technology, the ability to protect against tampering and data immutability to detect and prevent malicious activities and solve the data consistency problems faced by cooperative intrusion detection systems (CIDN). 

Figure 3 illustrates the organization of architectural blocks for a case study in an IDS blockchain system for hospitals and eHealth. The latter uses blockchain sparingly, primarily as a form of securely storing and extracting features or signatures, adding an additional verification step and making the storage of these signatures and features and the sharing of data secure.

\begin{figure}[ht]
\centerline{\includegraphics[width=1\columnwidth]{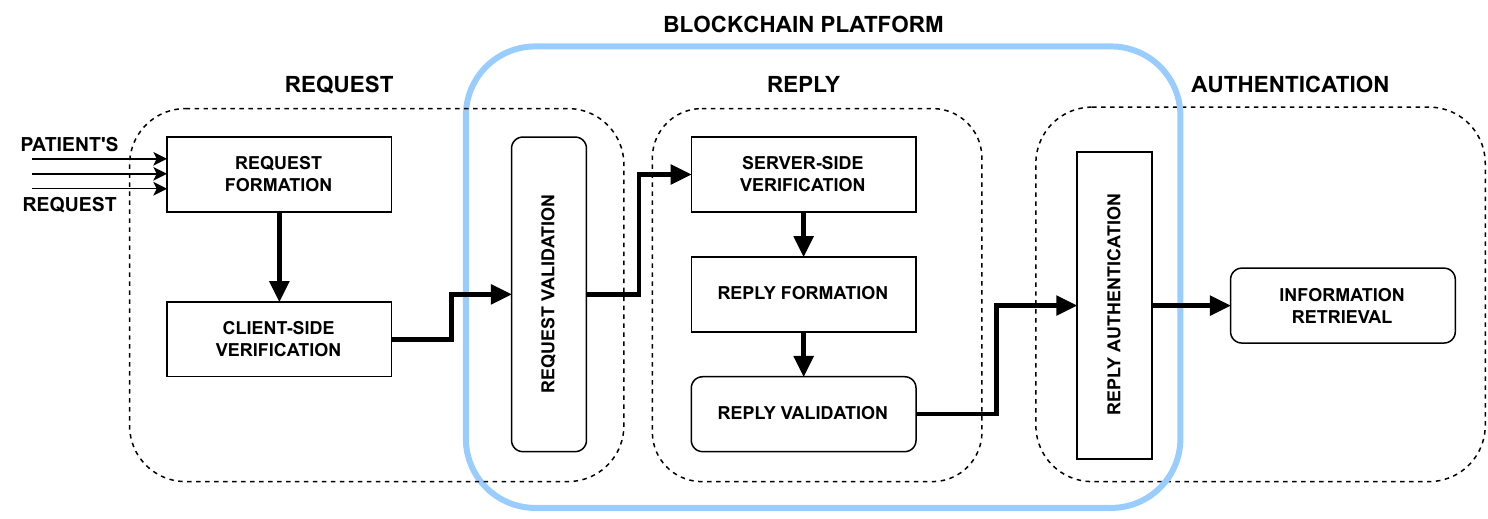}}
\centering
\caption{Building Blocks of Distributed Features in IDS with Blockchain (e-health application use case) \cite{ajayi2021blockchain}}
\label{fig:building_blocks_IDS}
\end{figure}

The idea of hybridization of systems from different disciplines and the value brought by the inclusion of multidisciplinary knowledge turns out to be a good indicator for the value and applicability of these solutions and architectures; therefore, our research also tries to expand and explore the scope of this approach in building systems for cyber protection and information security.

By adding to the recognition of user behaviors for anomaly detection as the most effective way of building IDS and through already known techniques for network traffic classification, this paper tries to expand the limits of integration by proposing a proof of concept for building a decentralized system architecture for the detection of cyber intrusions with the ability to credit, reward or evidence the ownership of the discoverer of cyber intrusions.
\vspace{-0.3cm}
\section{Intrusion detection systems (IDS)} \label{sec:sharing}

According to NIST, intrusion detection is the process of monitoring events and occurrences in a computer system or network and analyzing them for signs of potential incidents, which are violations or immediate threats of violation of computer security rules, acceptable usage rules or standard security practices. The breadth at which an IDPS operates creates a new division of IDPSs, which can be constrained within an application, a computing device, a network or within a multi-network infrastructure. In \cite{bhuyan2014network} various models and architectures of IDS are described in more detail that coincide and serve as a more practical illustration of these definitions so far.

To define cyber intrusion, we follow concepts defined by CERT/CC. While attacks and intrusions can be viewed from a number of perspectives, the most common are those of the attacker and the victim. Each view brings with it separate criteria for judging the success of the attack. Most of the time, it is sufficient to say that an intrusion has occurred when an attack is considered successful from the victim's perspective, i.e., the victim has experienced some loss, harm, or consequence. A successful attack or intrusion is enabled and achieved by the presence of a network vulnerability or vulnerability in the victim's system that is exploited by an attacker with an objective or goal in mind. In the following we use the word intrusion to mean such a successful attack.

On the other hand, an attack is considered unsuccessful from the attacker's point of view if none of its objectives are met; while the victim perceives the attack as unsuccessful if there are no consequences resulting from the attack. Unsuccessful attacks from an attacker's perspective can still have one or more consequences for a victim. The goal of intrusion detection is to accurately identify all true attacks and ignore all common situations. Like attacks, the process of detecting them can be defined from different perspectives. Intrusion detection can result from observing an attack in real-time or continuously or from knowing the results of an intrusion after it has occurred post-factum.

Figure 4 shows an illustrative architecture of the construction of an anomaly detection engine with layers for anomaly detection through classifiers. The construction of anomaly profiles through labeling in intrusion detection systems can then be subsequently managed through the software in the anomaly detection engine. The taxonomy and the anomaly detection engine complete the body of knowledge about the essentials and possible ways of building intrusion detection systems to date.

\begin{figure}[ht]
\centerline{\includegraphics[width=1\columnwidth]{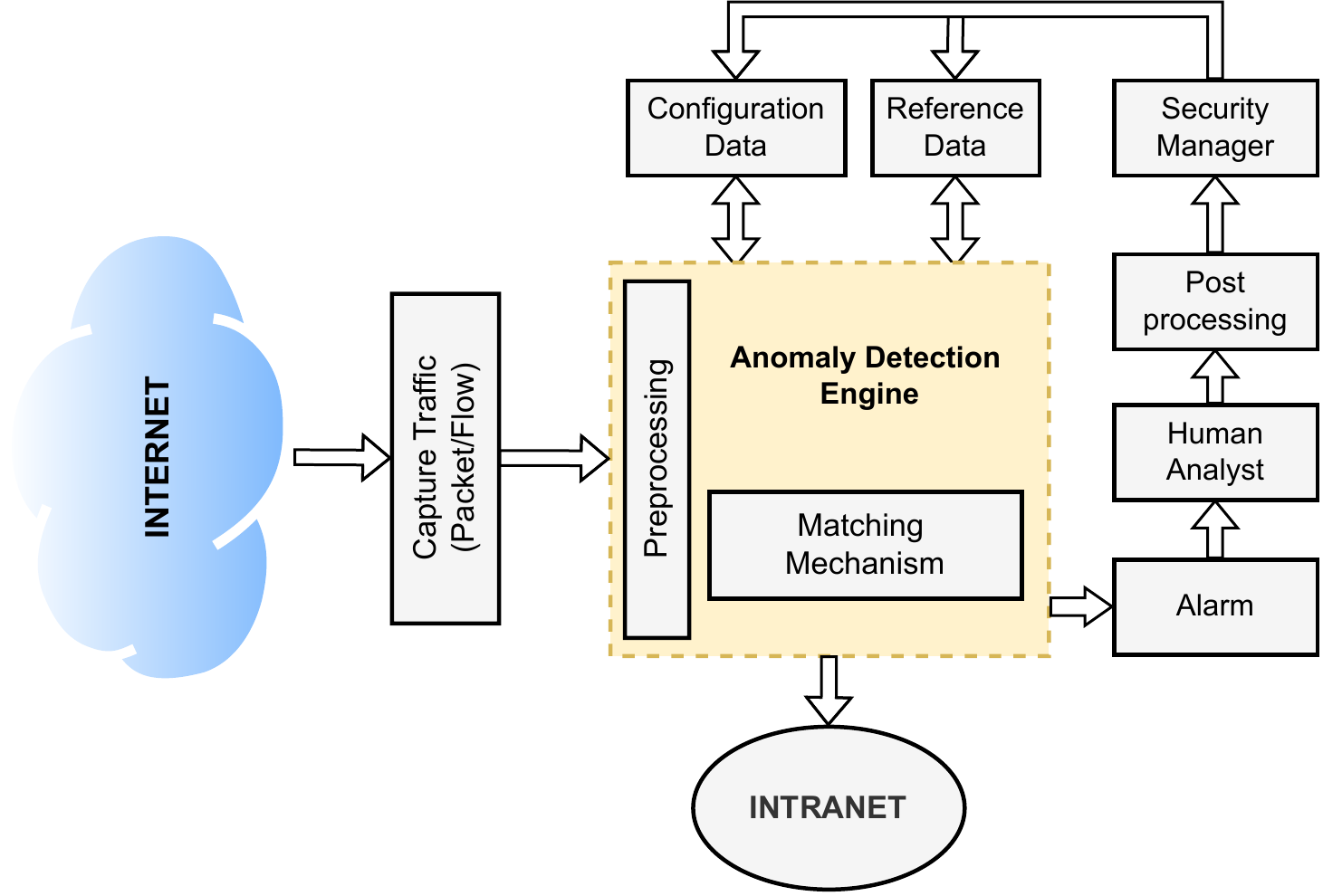}}
\centering
\caption{Architecture of an anomaly detection engine. Adapted from \cite{bhuyan2014network}}
\label{fig:architecture_integration}
\end{figure}
\vspace{-0.5cm}
\section{Network traffic analysis with ML} \label{sec:sharingschema}

Since we chose a supervised learning method and due to the specific architecture of this research, the proof of concept required at least one initial dataset and a model for training data at first before the system can operate fully autonomously and distributed. In that regard, we consider that it is worth discussing the details of traffic monitoring, the collected traffic data, the features of the data that are stored, the datasets that exist and their characteristics. With the increase of research in IDS, many different efforts have been made to publish relevant datasets for intrusion evaluation and modeling. The initial serious attempts to build datasets lead to the first reference datasets, i.e., datasets that have been used in many research papers and that have allowed the research community to draw many lessons in different fields of study. It is important to clarify at this point that we will consider as initial datasets only those datasets that include labels indicating whether attacks are present or not, due to the supervised learning nature of our ML algorithms.

From our review and researched reviews of existing datasets (e.g. in \cite{macia-fernandez2018ugr16}), it has been established that several requirements must be met in terms of the information that ought to be monitored in order to have results with acceptable reliability:
\begin{itemize}

\item {\em Network packets:} Datasets may include network and/or host properties. Network properties are usually summarized on a per-flow basis and include characteristics such as flow duration, timestamps, number of bytes and packets, flags, IP addresses and port numbers, average time between packets, etc. Regarding host properties, some examples used in datasets are: the number of failed communication establishment attempts or flags to indicate certain conditions, such as if super admin access is allowed or if the attacker is successfully authenticated.

\item {\em Real background traffic:} The dataset must contain real-world traffic. This is especially critical with background traffic, due to the fact that a synthetic generation of network traffic can lead to patterns and normal behavior that are not correct or do not have real-world leads.

\item {\em Up-to-date vulnerability traffic:} Datasets for IDS evaluation should contain real-world attack and intrusion cases updated with the latest intrusion techniques and tools.

\item {\em Labeling:} Different observations in datasets should ideally be accurately and precisely labeled as malicious or not. In case of attack observations, it is necessary to save the type of attack. This labeling process presents many difficulties with real data.

\item {\em Duration:} The data should extend over several test cycles (day or week), to make it possible to consider the cyclostationary evolution of traffic in the day/night period, as well as during weekdays/weekends and even for different months.

\item {\em Documentation:} It is paramount to have a comprehensive description of the data to understand its limitations and potentials. Some of the published datasets fail to provide sufficiently comprehensive information.

\item {\em Format:} Datasets typically include information in pcap or tcpdump, csv, or flow (netflow) format. The pcap format allows a more comprehensive evaluation of IDSs, mainly due to the fact that all information (payload) is included in the information. In the csv format, on the other hand, there is a pre-processing of the pcap information and an extraction of the relevant data for various features.

\end{itemize}

All of these factors are extremely important in choosing how we monitor, store, structure, and classify network traffic. This information can even make the difference between an effective intrusion detection system versus an ineffective one. In the following we will discuss how to define the data schema and the selection of test sets from initial datasets.

\subsection{Data schema}

During the selection of initial datasets, a set of recent datasets related to network traffic including MTA-KDD'19 \cite{letteri2020mta}, USB-IDS \cite{catillo2021usb}, CTU13 \cite{garcia2014empirical} and UGR16 \cite{macia-fernandez2018ugr16} were tested. Each of these datasets was reviewed against the above criteria and tested as potential initial datasets in local test environments. Due to the simplicity of the structuring, the high precision of SVM classification and the migrateable format, it has been decided to follow and use MTA-KDD'19 \cite{letteri2020mta} as the initial dataset and the initial format of data collection and structuring during traffic monitoring. In principle, any of the datasets could be used without any significant changes, however the format and size of the other data are more impractical. Integrating multiple datasets can be considered in future work. Finally, this dataset serves as part of the conceptual proof, since in a practical implementation it would have to be replaced with synthetic data or monitored and collected under suitable conditions depending on the applicability and the nature of the network where the classification would take place.
\vspace{-0.3cm}

\section{Decentralization}
\label{sec:collaborativeschema}
In order to discuss the application of decentralization in the context of IDSs, we will first present some of the main characteristics of blockchain technology, which today is considered the only widely applicable implementation of a fully decentralized, consistent and reliable data sharing system.

\subsection{Blockchain}

To reach to the proof of concept and for ease of illustration, a simplified version of the blockchain is implemented in this work and is used together with other components to build the prototype of the application. To briefly introduce decentralization, blockchain stores a series of sequential blocks where transactions and transaction information are placed. This blockchain is stored in a file locally on each node in json format. The choice of format was mainly for illustrative convenience and no better justification than that, otherwise there are better serialization formats that also save disk space for very old blockchains. Listing \ref{lst:block_chain} bellow presents a sample code of the resulting blockchain.

\begin{lstlisting}[language=Python,caption={Sample blockchain data},label={lst:block_chain}]
{
  "chain": [
    {
      "previousHash": "0",
      "timestamp": 1483228800000,
      "transactions": [],
      "nonce": 0,
      "hash": "cd1e9d208d0fa58d3e323758f9d59ed4f
      dc19e2292203cf18a9c34f2c032e182"
    },
    {
      "previousHash": "cd1e9d208d0fa58d3e323758f9d59ed4fdc19e
      2292203cf18a9c34f2c032e182",
      "timestamp": 1656867419168,
      "transactions": [
        {
          "fromAddress": null,
          "toAddress": "04c26dd943516d634b0da9781
          fadfad0f6902ccf51e6980167c477...",
          "amount": 10,
          "metadata": {},
          "timestamp": 1656867419168
        }
      ],
      "nonce": 2256202,
      "hash": "000008e6fa8ae04b3fd060028b3d8ebbb2c
      57ea1acf2375b9aee131990097aaf"
    },
    {
      "previousHash": "000008e6fa8ae04b3fd060028b3d8eb
      bb2c57ea1acf2375b9aee131990097aaf",
      "timestamp": 1656869513906,
      "transactions": [
        {
          "fromAddress": null,
          "toAddress": "04c26dd943516d634b0da9781fadfad0f6902ccf
          51e6980167c477b2e2c...",
          "amount": 10,
          "metadata": {
            "cyberNFT": "NewDiscovery"
          },
          "timestamp": 1656869513906
        }
      ],
      "nonce": 711633,
      "hash": "00000a55ff316a49eaf47bd1e0089c125c099
      af236ec3ac872f920f97f3c903b"
    },
  ],
  "difficulty": 5,
  "pendingTransactions": [],
  "miningReward": 10
}
\end{lstlisting}

Access and communication with the blockchain is done through endpoints implemented and exposed as Restful API with HTTPS for interaction between nodes using the same blockchain. It is worth noting that this blockchain implementation requires a private/public key pair, which is generated upon acceptance of the key parameter in all requests for the user's account private key. The establishment of difficulty and effectively time required to generate a new block and to place the transactions in the block is determined by the difficulty parameter. This parameter determines how many finite digits in the hash value of the last block must be guessed to generate a block. The larger this number is, the more time is required to generate new blocks.

For clarity and a more comparable presentation of the values, a logarithmic diagram is given in Figure 5, where the time axis is presented in logarithmic units. It is clear that the growth is extremely fast, almost exponential, and this flow is a result of the irreversible nature of hash functions and the effort to find or reverse the value through brute force techniques. As explained in a mathematical representation later, this has a significantly costly impact to the overall speed of the IDS system.

\begin{figure}[ht]
\centerline{\includegraphics[width=1\columnwidth]{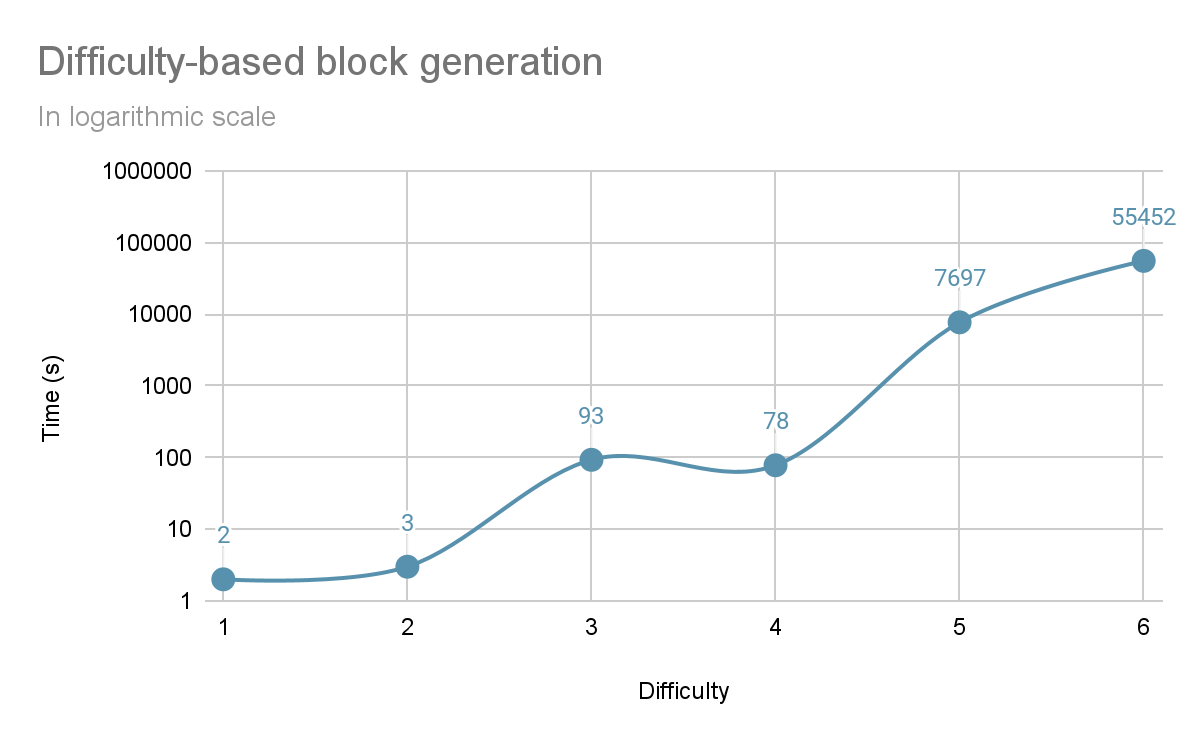}}
\centering
\caption{Logarithmic relationship of block generation time against difficulty level}
\label{fig:logarithmic_relationship}
\end{figure}

Since this implementation is effectively in accord with blockchain specifications and has accounts (wallets) and transactions, then it is understandable that even in our implementation it will be possible to transfer the rewards obtained from the generation of a block between accounts, similar to existing cryptocurrency implementations.

\subsection{CyberNFTs} \label{sec:multiformschema}

From the discussion in the next section 6.1 it is more than clear that the integration of NFTs (Non-fungible token) in the above implementation is quite intuitive and easy. Adding a new metadata field to each transaction will suffice to hold and share additional data as long as the maximum allowed block size is not exceeded \cite{wilson2022prospecting}. We can take this concept and redefine it in the field of cyber security and cyber intrusion detection systems to introduce a reward-driven autonomous system.

A cyber non-fungible token, or CyberNFT, is defined as a unique hash value derived from an exploit/intrusion signature (in signature-based IDS) or information, which is placed as metadata in mined blockchain transactions to mark ownership by the account that first mined this information in the blockchain or whose it was last transferred to the blockchain by the owner. In principle, CyberNFTs are a way of attributing the ownership of the discoveries of new cyber intrusions in the form of the signature with which they were discovered.

The proof of concept uses a local web server to show all existing transactions on the blockchain, the relevant cyberNFTs and the rewards that the accounts have earned in the form of tokens in exchange for creating blocks. While Algorithm 1 shows an approach of implementing the storage of cyberNFT information as metadata in transactions of this nature. It can be seen that, in principle, in addition to cyberNFT, data can be placed in the metadata according to the user's specification. CyberNFTs and this way of crediting ownership can even be extended to an entire market, where discovery tokens are sold or bought and intrusion detection is self-sustained and self-rewarded. In addition, the tokens also serve as anomaly signatures, which can then be used in data modeling with ML.

While the infrastructural components of the proposed approach are well-known, it is important to acknowledge that, to the best of our knowledge, the specific application and approach of integrating these foundational; components represents a novel concept in addressing the challenges posed by traditional centralized intrusion detection systems. As a result, direct comparisons to existing research may be difficult.





\begin{algorithm}
\caption{Implementation of CyberNFT Information Storage}
\begin{algorithmic}[1]
\State \textbf{Class} Transaction
\Procedure{Constructor}{from, to, amount, metadata = \{\}}
    \State this.from $\gets$ from
    \State this.to $\gets$ to
    \State this.amount $\gets$ amount
    \State this.metadata $\gets$ metadata
    \State this.timestamp $\gets$ Date.now()
\EndProcedure
\Procedure{mineOneTransaction}{miningRewardAddress, metadata}
    \State cyberNFTx $\gets$ new Transaction(null, miningRewardAddress, this.miningReward, metadata)
    \State Add cyberNFTx to this.pendingTransactions
    \State Create a new instance of 'Block' with the following parameters:
    \begin{itemize}
        \item Current timestamp using 'Date.now()'
        \item The 'upcomingTransactions' array
        \item The hash of the latest block in the chain
    \end{itemize}
    \State Call the 'mineBlock' method on the new 'Block' instance with the 'difficulty' parameter
    \State Add the mined 'Block' to the 'chain' array
    \State Reset the 'upcomingTransactions' array to an empty array
\EndProcedure
\end{algorithmic}
\end{algorithm}

\section{Case Study – Autonom}\label{sec:tag}

To further the analysis of the interaction between these latest state-of-the-art machine learning (ML) and blockchain technologies in the field of cyber security, we have simulated such an autonomous system, in which we can be identify very easily applicability challenges alongside existing, outdated and conventional systems used today in centralized or semi-centralized on-premise and cloud architectures.

In the previous parts, we discussed the features of this system implemented in isolation, so we discussed the capabilities for it to:
\begin{itemize}
    \item monitor local network traffic
    \item utilize ML classification models
    \item classify and store traffic in a decentralized manner on the blockchain
    \item accentuate and reward the intrusion detection peer with a cyberNFT
\end{itemize}

\begin{figure}[ht]
\centerline{\includegraphics[width=1\columnwidth]{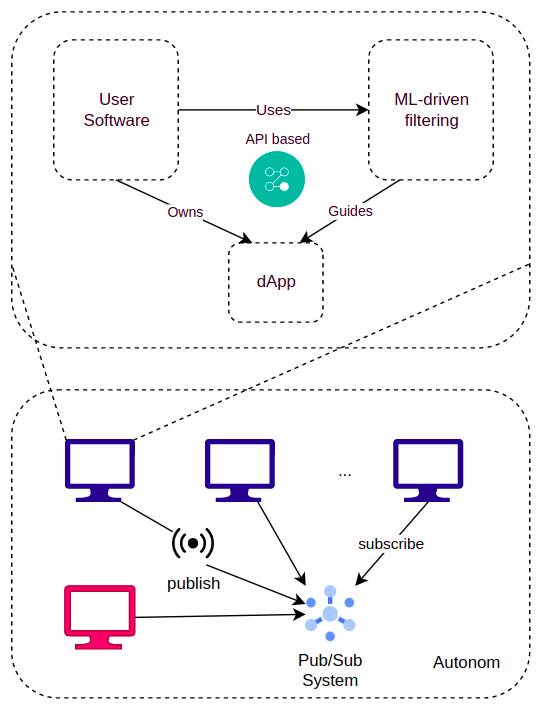}}
\centering
\caption{Interaction between devices in the autonomous network}
\label{fig:autonomous_network}
\end{figure}

Figure \ref{fig:autonomous_network} illustrates the interaction of these components in a holistic manner through publish/subscribe mediation; in the following section we will clarify further a use case, which simulates a utilization scenario of this solution.
\begin{enumerate}
    \item The user visits different pages and uses network traffic
    \item The user generates traffic that is classified as abnormal
    \item The IDS performs a preliminary classification on the local device
    \item IDS classifies the behavior as suspicious communication and generates the report/notification
    \item The IDS updates the blockchain with the detected anomaly detection generates a cyberNFT
    \item The user transfers the cyberNFT to an existing or new account for that node
    \item After the blockchain is updated, the blockchain account associated with that IDS receives a reward cyberNFT
    \item The IDS after a certain amount of time sends the collected data to a subset of nodes that can update the model
    \item Nodes update the model with the latest information
    \item The user visits other sites
\end{enumerate}

These steps illustrate all the main use cases of this hybrid blockchain and ML IDS system. To enable communication between nodes, however, a subscription/publication (or publish/subscribe) system is needed, which are very common in larger infrastructures or in the implementation of architecture with microservices between distributed devices/processes. The implementation of this system is mainly necessary for nodes to notify each other in case of blockchain update and to avoid servers altogether.

In the proposed decentralized intrusion detection system (IDS), we employ a peer-to-peer (P2P) network and leverage blockchain technology for data storage. It is crucial to address the inherent costly delay disadvantage associated with such systems. To provide a mathematical representation of the proposed system, we consider several factors influencing node delay, propagation delay, and block generation delays. Subsequently, we define the Mean Time to Detect (MTTD) and Mean Time to React (MTTR), using queuing theory to model the performance of the P2P network within this IDS context \cite{shortle2018fundamentals}.

The MTTD is influenced by factors such as the data processing capability of each node, the network propagation delay, and the efficiency of the detection algorithm. Within a weighted graph of this network, where the weight of each node represents its processing power, we can calculate the average MTTD as follows:

\begin{equation}
\label{eq:1}
MTTD=\mathbb{E}\left(\frac{1}{\sum(\mu_i)}\right)
\end{equation}

In Equation \ref{eq:1}, $\mathbb{E}$ denotes the expected value (to account for the uncertainty in the service rates of the nodes) \cite{shortle2018fundamentals}, and $\mu_i $ represents the service rate of node i, which is determined by factors such as processing power, detection algorithm efficiency, and workload \cite{butun2014survey}. The service rate ($\mu$) can be determined by analyzing the network's processing power, the efficiency of the intrusion detection algorithm, and other factors affecting the node's ability to process incoming requests. The arrival rate ($\lambda$) can be estimated by analyzing the frequency of incoming packets or requests, which could be influenced by factors such as network traffic, the number of active nodes, and network topology.

MTTR is influenced by the time taken to add a new block to the blockchain and the time required to propagate information across the network. The block generation time (T) is a crucial parameter that impacts the overall network performance. It can be estimated using the formula:
\begin{equation}
\label{eq:2}
T = \frac{D}{H}
\end{equation}

where in Equation \ref{eq:2}:

\textbf{T} is the average time to generate a block

\textbf{D} is the current difficulty of the network

\textbf{H} is the total processing power (hash rate)
\\A longer block generation time can be attributed to factors such as computational complexity, network hash rate, variability, and difficulty adjustments \cite{lin2017survey}.

Let’s denote the arrival rate ($\lambda$), which can be estimated by analyzing the frequency of incoming packets or requests, and could be influenced by factors such as network traffic, the number of active nodes, and network topology. The time taken to propagate information across the network can be estimated using queuing theory. In the context of a P2P network, the expected waiting time in the queue ($W_q$) can be computed using the M/M/1 queue formula \cite{shortle2018fundamentals}, applicable under the condition that the traffic intensity ($\rho = \lambda/\mu$) is less than 1:
\begin{equation}
\label{eq:3}
W_q = \frac{\lambda}{\mu\cdot\left(\mu - \lambda\right)}
\end{equation}

In Equation \ref{eq:3}, $\lambda$ is the arrival rate of packets, and $\mu$ is the overall service rate. The expected time to react (MTTR) in the system can be estimated by considering both the waiting time in the queue ($W_q$) and the block generation time (T), which is presented in Equation \ref{eq:4}

\begin{equation}
\label{eq:4}
MTTR = W_q + T
\end{equation}

It is important to note that this is a simplified and illustrative mathematical representation, and real-world implementations may require more complex models and simulations for specific network configurations \cite{sun2011survey}. While the M/M/1 queuing model assumes a single server and exponentially distributed arrival and service times, it may not accurately represent the P2P network and decentralized IDS. Other queuing models such as M/M/c or M/G/1 could be considered in a P2P network. Nonetheless, this representation serves as a foundation for further analysis and development in the domain of decentralized intrusion detection systems and highlights the impact that block generation time has on the overall MTTD and MTTR.

This proof of concept and the effectiveness of the decentralization of security mechanisms evaluate the expansion possibility more closely in terms of its practical application. The main challenges presented in this integration are not related to the architectural choises per se, but mainly related to the limitations of the blockchain technology itself \cite{golosova2018advantages} like:
\begin{itemize}
    \item Gradual high power consumption,
    \item Limited block size,
    \item The proportional decrease in speed with the growth of the blockchain,
    \item Blockchain security threats,
    \item Scalability, etc.
\end{itemize}

In addition to the challenges in implementing a decentralized database, the management and construction of classification models in terms of memory and processing resources must also be done, taking into account that the amount of traffic on many distributed nodes grows extremely fast, much faster than any single node or cluster of nodes can handle. In this situation, a solution is required for building the joint classification model through distributed programming or by choosing nodes from which the traffic is used for building the model unilaterally, prioritizing those nodes in closer proximity to where anomalies were found in preliminary classifications.

\subsection{Integration}

Figure 6 presents a general scheme of the integration of all components in a network of devices, where each node monitors its local traffic and distributes intrusion signature information to all other nodes in real time, as well as propagating changes through to the pub/sub network that mediates in the middle. The user software, in addition to traffic monitoring, can also perform other functions such as storing monitoring history, scanning files, interfacing and processing information from dApp, blockchain and cyberNFTs, etc.

Traffic classification with ML, specifically with Support Vector Machines algorithms is done on selected nodes, which have the memory and processing capacity to run these models with ease. In the case of a network with insufficient components, it will be necessary to establish several nodes for this purpose and inevitably bring back elements of centralization to the network due to the lack of resources. All classifications of all nodes are weighted and measured according to consensus protocols, similar to those of the blockchain itself. In the case of infected nodes, the software program must have a mechanism for breaking communication from abnormal traffic without user response, to avoid spreading the intrusion to the entire network.

\subsection{Distributed operation (pub/sub)} \label{sec:coltag}

The publish/subscribe (or pub/sub) paradigm is a widely used model for interfacing applications/processes in a distributed environment. Many existing pub/sub systems are based on default entities or predefined non-dynamic nodes, and therefore are able to exploit multicast technologies to provide scalability and availability. An emerging alternative to these allows information consumers to search for events based on the content of published events. This is usually organized by dynamically creating/deleting subscription and publishing channels.

This is best illustrated in Figure 7, where the information space for subscription/publication is presented alongside their users. In our case, all members of the autonomous network subscribe to the same information space, which, depending on the situation, behaves as a space for broadcasting or a space for multicasting.

\begin{figure}[ht]
\centerline{\includegraphics[width=1\columnwidth]{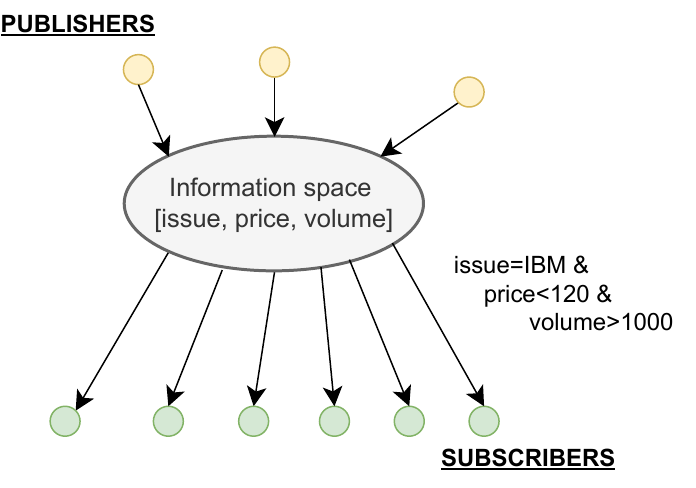}}
\centering
\caption{Subscription/publish information space \cite{banavar1999efficient}}
\label{fig:subscription_publish}
\end{figure}

\vspace{-0.2cm}
\section{Discussion}

Web3 technologies challenge the virtual world of finance, data, intellectual property and the like on an ongoing basis. The return to the decentralization of data storage and distribution has prompted the creation of entirely new ways of doing transactions, information processing, value exchange, art, etc. Yet, the study of interaction and application of this decentralized way for the implementation of collaborative intrusion detection systems has not been sufficiently addressed at the implementation level.

Since the creation of cyber intrusion detection systems it is evident that there has been much progress in the integration and incorporation of machine learning into these systems. We can even say that the old ways of detecting anomalies and signatures have been almost completely replaced by algorithms of machine learning (ML), deep learning (Deep Learning), neural networks (CNN) and variants of anomaly detection engines. The outdated ways through regular classification engines are almost no longer in use. On the other hand, the decentralization of IDS systems is an effort, which has turned out to be less researched and more challenging for the scientific community.

Alongside the anonymity, immutability, availability and other inherent benefits, blockchain also brings a series of difficulties and obstacles, which require a hefty amount of work and research to be applicable in the emergency context to which intrusion detection systems operate. Problems of cost, scalability, block size limitations and the like present a wider gap and divide from the level of applicability a decentralized IDS requires. However, our case study verifies once again that under controlled conditions if not the entire infrastructure, at least a partial decentralization in terms of ownership attribution, preliminary information and reports can is feasible. The cross-node generation of classification models and the publish/subscribe system implementation require further study to enable a true decentralized architecture, instead of the quasidecentralization that their limitations impose.

\begin{table}[htb]
\centering
\adjustbox{max width=\columnwidth}{
\begin{tabular}{
>{\columncolor[HTML]{F3F3F3}}l 
>{\columncolor[HTML]{F3F3F3}}l 
>{\columncolor[HTML]{F3F3F3}}l 
>{\columncolor[HTML]{F3F3F3}}l l}
\cellcolor[HTML]{D9D9D9}\textbf{Features} & \cellcolor[HTML]{D9D9D9}\textbf{Host-based IDS} & \cellcolor[HTML]{D9D9D9}\textbf{Server-based IDS} & \cellcolor[HTML]{D9D9D9}\textbf{Serverless IDS} &  \\
Cost                                      & Low                                             & Medium                                            & Medium                                          &  \\
Speed                                     & Medium                                          & Medium                                            & Low                                             &  \\
Reliability                               & Low                                             & Medium                                            & Medium                                          &  \\
Availability                              & Medium                                          & Medium                                            & Medium                                          &  \\
Scalability                               & Medium                                          & Medium                                            & Low                                             &  \\
Precision                                 & Low                                             & Medium                                            & Medium                                          &  \\
Transparency                              & Low                                             & Low                                               & Medium                                          & 
\end{tabular}}
\caption{Comparison of features between host-based, server-based and serverless IDS systems}
\label{tbl:feature_comparison}
\end{table}

Table \ref{tbl:feature_comparison} compares cost, speed, reliability, availability, scalability, precision and transparency between three categories of IDS systems, separated by locality. We name an IDS host-based, when the IDS is installed and isolated in a system with little to no communication outside the software. An IDS is considered server-based, when there is some data or information stored in a server outside the installed IDS software. With serverless IDS we categorize a new type of IDS system, which shares and distributes information between peers, but without the inter-mediation of servers and through decentralized collaboration.

As discussed above, the integration of blockchain in IDS systems would increase the reliability, availability and quality of information shared between nodes in a collaborative network, however, with the growth of the network, the cost and speed of this network would drop significantly according to preliminary results and as a result reduce its effectivity.

\section{Conclusion}

The paper presents an analytical and comparative study of the interaction between the latest state-of-the-art Web3 technologies in the context of information security and expands on their practical application in this field. For illustration and proof of concept, an early prototype for detecting cyber intrusions using machine learning modeling as a decentralized and distributed application (dApp) has been implemented and analyzed. The prototype highlights the comparisons and differences between this decentralized way of implementation in Web3 versus the common Web2-compliant centralized implementation.

A completely decentralized implementation of an IDS system has not reached the level of service maturity to compete with current solutions, except in some isolated or context-specific environments where we have enough room for pre-defining and decentralized network nodes.

Consequently, the case study shows the shortcomings of this approach and elaborates the possibilities for further study towards a solution which mitigates the hindering qualities of blockchain, the features of the centralized nature of the subscription/publication systems and the processing cost of building classification models.

The research so far and the results of this paper define the main obstacles and pave the way for the complete decentralization of the prototype we have developed. In future works we will explore more efficient and decentralized ways of publish/subscribe concept, we will explore the possibility of improving classifiers using multi-level neural network classification models and we will evaluate the experimentation and testing of the developed prototype in hybrid cloud/on-premise ecosystems to observe if we have comparable results to the actual ones on unique devices with explicit communication. In addition, the paper highlights and leaves room for further discussion of the main obstacles and limitations that the build of blockchain itself brings in terms of scalability. This review creates space for a new implementation of blockchain, which is more compatible with the technical constraints of to-date nodes for detecting cyber intrusions.

\printbibliography

\end{document}